\documentclass[showpacs,epsf,acs,onecolumn]{revtex4-1}
\usepackage{float}
\usepackage{color,graphicx}
\usepackage{hyperref}
\begin{document}

\title{Prominent metallic surface conduction  and the singular magnetic response of  topological Dirac fermion in three-dimensional topological insulator Bi$_{1.5}$Sb$_{0.5}$Te$_{1.7}$Se$_{1.3}$}

\author{Prithwish Dutta$^{1,2}$, Arnab Pariari$^{1}$, and Prabhat Mandal$^{1}$}
\email{prabhat.mandal@saha.ac.in}
\address{$^1$Saha Institute of Nuclear Physics, HBNI, 1/AF Bidhannagar, Calcutta 700 064, India}
\address{$^2$Goverment General Degree College, Singur, Hooghly 712409, India}
\date{\today}
\begin{abstract}
{\large We report semiconductor to metal-like crossover in temperature dependence of resistivity ($\rho$) due to the switching of charge transport from bulk to surface channel in three-dimensional topological insulator  Bi$_{1.5}$Sb$_{0.5}$Te$_{1.7}$Se$_{1.3}$.  Unlike earlier studies, a much sharper drop in $\rho$($T$) is observed below the crossover temperature due to the dominant surface conduction.  Remarkably, the resistivity of the conducting surface channel follows a rarely observable $T^2$ dependence at low temperature as predicted theoretically for a two-dimensional Fermi liquid system. The field dependence of magnetization shows  a cusp-like paramagnetic peak in the susceptibility ($\chi$) at zero field over the  diamagnetic background. The peak is found to be robust against temperature and  decays linearly with field from its zero-field value. This unique behavior of $\chi$ is associated with the spin-momentum locked topological surface state of  Bi$_{1.5}$Sb$_{0.5}$Te$_{1.7}$Se$_{1.3}$. The reconstruction of surface state with time is clearly reflected through the reduction of peak height with the age of the sample.}
\end{abstract}
\maketitle

Topology protected electronic band structure is of current interest in condensed matter and material science research. Three-dimensional (3D) topological insulator has been established as an important member of this family since its first realization in Bi$_{1-x}$Sb$_{x}$\cite{Heish} and in binary compounds such as Bi$_{2}$Se$_{3}$\cite{Xia}, Bi$_{2}$Te$_{3}$, Sb$_{2}$Te$_{3}$\cite{HD}, etc. This class of `insulators', due to strong spin-orbit coupling, hosts a metallic surface state with unique physical properties \cite{hasan}. The spin-momentum locked surface state, which is protected by the time-reversal symmetry, makes the electronic transport robust against inelastic backscattering\cite{hasan, qi}. Theoretical investigations have predicted them as potential candidates for applications in spintronics, quantum computation, etc\cite{qi}. However, the above mentioned topological insulators reported to exhibit high bulk conductivity, instead of a gap between the valance and conduction band \cite{eto, qu, dutta, Hor}. It is now well documented that crystal defects - namely antisite defects and vacancies, which introduce large  residual carriers, are the main reasons behind such high conductivity of the bulk  \cite{Hor,Taskin, Ren}. As a result, the transport response is dominated by the bulk rather than the surface. The misplacement of Se atom from its lattice position causes an increase in negative carrier concentration, which can be as high as $\sim$ 10$^{19}$ cm$^{-3}$, whereas  the antisite defect of Bi and Sb  produces holes \cite{Ren}. Thus by replacing Bi with Sb and Se with Te in a proper ratio, one can balance the opposite type of carriers and minimize the conductivity. In order to achieve the  higher figure of merit, efforts have been made to decrease the carrier density in the bulk by using ternary and quaternary compounds of Bi, Sb, Te and Se \cite{Ren, shekhar}.   Teramoto  and Takayanagi,  in their work, reported  that high resistivity can be obtained for compounds with chemical formula Bi$_{2-x}$Sb$_{x}$Te$_{3-y}$Se$_{y}$ (BSTS)  for a certain linear relation between $x$ and $y$ \cite{Teramoto}. Ren \emph{et al.}  reinvestigated the electronic and structural phase diagram to achieve the intrinsic topological insulating state in these compounds and prescribed that the highest resistivity can be realised close to composition Bi$_{1.5}$Sb$_{0.5}$Te$_{1.7}$Se$_{1.3}$  \cite{Ren}.  Since then BSTS has been studied by various groups through transport  \cite{Ren, Ko, Fran, Taskin, Segawa, Lee, Pan, Hsiung} and spectroscopic measurements\cite{Arakane, Kim, Ko, Post, Fran}. \\

Topological insulators are characterised by their symmetry protected  novel surface state properties. However, to observe  high quality surface state property, one needs to electronically decouple the surface from the bulk. This can be achieved, by making the bulk highly insulating. We have grown high quality single crystals of topological insulators with composition Bi$_{1.5}$Sb$_{0.5}$Te$_{1.7}$Se$_{1.3}$. The bulk resistivity of some of the crystals used in the present study is about an order of magnitude larger than  earlier reports \cite{Ren, Ko, Fran, Taskin, Segawa, Lee, Pan, Hsiung}. Notably, resistivity  at low temperature decreases quadratically with temperature which is a clear evidence of Fermi liquid behavior of the surface state.  Furthermore, the field dependence of susceptibility ($\chi$) exhibits an unusual paramagnetic peak at zero field as an evidence of the spin-momentum locked Dirac cone surface state.\\

{\large \textbf{Results}}\\
\textbf{Temperature dependence of resistivity of Bi$_{1.5}$Sb$_{0.5}$Te$_{1.7}$Se$_{1.3}$  single crystals.} The  temperature dependence of resistivity ($\rho_{xx}$) for three  freshly cleaved Bi$_{1.5}$Sb$_{0.5}$Te$_{1.7}$Se$_{1.3}$  single crystals (S1,S2,S3) are presented in Fig. 1(a).  Two slightly different techniques were used to prepare these single crystals. Crystal S3  was prepared by method I while crystals S1 and S2 were prepared by method II. The details of preparation are described in the Method section.  With decreasing temperature,  resistivity initially increases rapidly and then either decreases or tends to saturate. Overall, $\rho_{xx}$ is  larger for the crystals S1 and S2 as compared to S3. For example, $\rho_{xx}$ for sample S2 is about 0.7 $\Omega$ cm at 300 K and it increases with decreasing temperature and reaches a maximum of 42 $\Omega$ cm at $\sim$ 40 K. On the other hand, $\rho_{xx}$ at 300 K is an order of magnitude smaller and $\rho_{xx}$($T$) shows a saturation-like behavior at low temperature  for sample S3. Though the crystals S1 and S2 were prepared by the same method, the value and $T$ dependence of $\rho_{xx}$ for these crystals are slightly different. Ren et al. also reported sample dependence of  $\rho_{xx}$ which they attributed to small variation in defect concentration in their samples \cite{Ren}. For a given nominal composition, they observed $\rho_{xx}$ to vary by a factor of 3 \cite{Ren}. Both the value and $T$ dependence of $\rho_{xx}$ for sample S3 are similar to earlier reports \cite{Ren, Taskin}.  In a usual electrical transport measurement, as the bulk and surface resistance are in  parallel configuration, they contribute simultaneously to conductivity.  However, their relative contribution is sensitive to temperature. Due to the two-dimensional (2D) nature of the surface state, the cross section area of the conducting channel is extremely small. For this reason, the surface resistance can be significantly larger as compared to semiconducting bulk specially at high temperatures. Thus, the charge conduction at high temperature is dominated by the bulk.  With decrease in temperature, the bulk resistance increases exponentially whereas the surface resistance decreases. This kind of temperature dependence of bulk and surface resistance results in a semiconductor to metal-like crossover at low temperature below which the charge transport is dominated by the surface\cite{Taskin,Lee,  Fran, Ko, Pan}. For both  technological application and basic research,  it is desirable that the bulk conductivity should be very small while the surface conductivity should be high. Several attempts have been made to achieve low bulk conductivity.  Most of the reported data on single crystals show that $\rho_{xx}$ exhibits the expected semiconducting behavior at high temperature. However,  at low temperature,  $\rho_{xx}$ becomes either  $T$-independent or shows a very weak temperature dependence, i.e., the system is barely metallic. Also,  the maximum value of resistivity is few $\Omega$ cm only. $\rho_{xx}$ for S1 and S2 samples at semiconductor to metal-like crossover temperature is an order of magnitude larger than those reported by several groups\cite{Ren, Taskin, Lee, Pan} and falls rapidly with decreasing temperature down to 2 K. To the best of our knowledge, such a strong decrease in surface-state resistance with decreasing temperature has not been observed so far. \\
\begin{figure*}
\includegraphics[width=\textwidth]{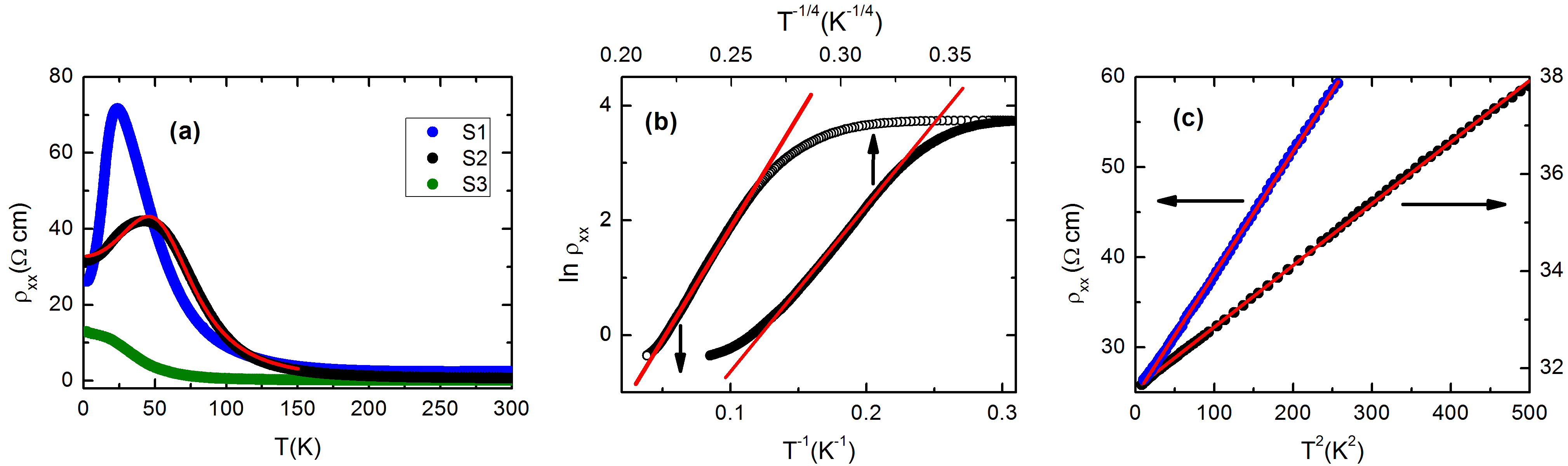}
\caption{(a) The temperature dependence of resistivity ($\rho_{xx}$) of Bi$_{1.5}$Sb$_{0.5}$Te$_{1.7}$Se$_{1.3}$ single crystals collected from two different batches. Samples S1 and S2 were prepared by method I and S3 was prepared by method II. The red line is the fit to the experimental data with equation $\rho_{xx}$$=$$[ 1/\rho_{s} + 1/\rho_{b}]^{-1}]$, where $\rho_{s}$($=a + bT^{2}$) and $\rho_{b}$($=\rho_{0}\exp(\frac{T_{0}}{T})^{1/4}$) are respectively surface and bulk resistivity.  (b) $\ln \rho_{xx}$ as a function of 1/$T$ (bottom axis) for activation behavior and as a function of $T^{-1/4}$(top axis) for variable-range-hopping behaviour for S2 crystal. (c) $\rho_{xx}$ is plotted against $T^{2}$ below 23 K. The linearity can be traced  up to 16 K and 21 K for S1 and S2 crystals respectively, confirming the Fermi liquid behavior of the surface state.}\label{Fig.1}
\end{figure*}

In order to understand the charge conduction mechanism, the temperature dependence of resistivity has been analysed. At high temperature, the insulating behavior of $\rho_{xx}$ can be fitted with the Arrhenius equation $\rho_{xx}$$=$$\rho_{1}\exp(\frac{\Delta}{k_{B}T})$.  From the linear behavior of ln$\rho_{xx}$ vs 1/$T$ plot as shown in Fig. 1(b) for sample S2, we have deduced the value of activation energy, $\Delta$$\simeq$40 meV. This value of activation energy is comparable to earlier reports \cite{Ren, Pan, Ko}. However, in the low-temperature region, $\rho_{xx}$ cannot be fitted with the activated type conduction.  We have tried to fit the resistivity data with 3D  Mott variable-range-hopping equation, $\rho_{xx}$$=$$\rho_{0}\exp(\frac{T_{0}}{T})^{1/4}$, which has been shown in Fig. 1(b). In this case, a better fitting is obtained with $T_{0}$ $\sim 5\times10^{6}$ K. The observed value of $T_0$ is quite large. $\rho_{xx}$($T$) for S1 and S3 samples exhibits qualitative similar behaviour.  Thus bulk conductivity shows activation behavior at high temperature but it switches over to 3D variable range hopping  at low temperature.  Crossover from activation to variable range hopping has also been reported earlier \cite{Ren, Pan}. In most of the cases, activation conductivity is observed in a narrow temperature range. In spite of much larger energy gap for some compositions, the bulk resistivity at low temperature is  smaller than the present samples \cite{Pan}. This is possibly due to the smaller value of $T_0$ (like the present S3 sample) which determines the bulk conductivity at low temperature.\\

For S1 and S2 crystals, the low-temperature resistivity below the  peak, which is dominated by the surface state, shows an upward curvature, suggesting a superlinear temperature dependence of $\rho_{xx}$ in the metallic region. We observe that $\rho_{xx}$ below 21 K for sample S2 can be fitted well with the equation $\rho_{s}$$=$${a + bT^{n}}$ for $n$=2. Figure 1(c) shows $\rho_{xx}$ vs $T^2$ plot. It is clear from the figure that $\rho_{xx}$ versus $T^2$ is strictly linear up to about 21 K. For crystal S1 also, $\rho_{xx}$ exhibits $T^2$ dependence below 16 K.  The $T^2$ dependence of resistivity at low temperature is the manifestation of Fermi liquid behavior of the surface carries.  Electron-electron scattering is known to give rise a $T^2$-dependent term in resistivity. The  $T^2$ dependence of resistivity is observed in various strongly correlated electron systems  such as heavy fermion metals, organic conductors and transition metal oxides. Though the $T^2$ behavior of resistivity is observed in several 3D Fermi liquid systems, it is rarely observed in 2D Fermi liquid systems. After creating a lateral magnetic superlattice, $T^2$-dependent resistivity due to electron-electron Umklapp scattering has been observed in 2D electron gas at GaAs/AlGaAs heterointerface \cite{mess,kato}. The temperature dependence of relaxation time in electronic transport can differ from that of quasiparticle.  Also, the relaxation time ($\tau$)  of quasiparticle of a Fermi liquid system depends on the dimensionality.  For example, $\tau^{-1}$$\propto$ $T^2$ in the 3D case whereas $\tau^{-1}$$\propto$  $T^2$ln($E_F/k_BT$) in the 2D case, where $E_F$ is the Fermi energy \cite{uryu}. However, the resistivity is proportional to $T^2$ in 3D as well as 2D systems \cite{uryu}.\\

\textbf{Magnetoresistance and weak anti-localization (WAL) effect in Bi$_{1.5}$Sb$_{0.5}$Te$_{1.7}$Se$_{1.3}$  single crystals.} The magnetoresistance (MR) defined as [$\rho_{xx}(B)-\rho_{xx}(0)]/\rho_{xx}(0)$ has been measured in the field ($B$) range 0 to 9 T. The field dependence of MR  at different temperatures is  shown in Fig. 2(a) for the S2 sample as a representative.  At low temperature,  MR increases almost linearly in the high field region. The nonlinearity of MR at low fields is due to the weak anti-localization  effect \cite{Lee, Chen}. With the application of magnetic field, the time reversal symmetry breaks down and a gap opens in the  surface state. Hence, the rate of change of magnetoresistance, which is maximum in the limit $B$ $\rightarrow$ 0, decreases with the increase in field strength. The  well known Hikami-Larkin-Nagaoka (HLN) formula used to describe the WAL effect is given by \cite{hln},
\begin{equation}
\bigtriangleup G(B) = -\frac{\alpha e^{2}}{2\pi^2 \hbar}[\psi(\frac{1}{2} +\frac{B_{\phi}}{B}) - \ln(\frac{B_{\phi}}{B})].
\end{equation}

Here, the magneto-conductivity ($\Delta G$) is defined as $\Delta G$=[$G$($B$) - $G$(0)], $G$ is surface channel conductivity, the parameter $\alpha$ represents the number of conduction channels (i.e., $\alpha$$=$0.5 for each conducting channel), $B_{\phi}$$=$$\frac{\hbar}{4el^{2}_{\phi}}$ and $l_{\phi}$ is the dephasing length. The WAL effect is solely a surface property and $l_{\phi}$ is the measure of its strength. The quadratic field dependence of magnetoresistance due to bulk is absent at low temperature. Thus, MR in the low temperature region originates from the conducting surface. The total electrical conductivity ($\sigma_{t}$) of  the single crystal is the sum of surface channel conductivity $G$ and bulk conductivity ($\sigma_{b}$), $\sigma_{t}$=$\sigma_{b}$ + $G/t$, where $t$ is the thickness of the sample. At low temperature, as the conductivity of bulk is very small compared to surface, $\sigma_{t}$$\simeq$$G/t$ is a good approximation.  As $\rho_{xx}$ is a tensor quantity in presence of magnetic field, $\sigma_{t}$  has been calculated using the relation $\sigma_{t}$=$\rho_{xx}$/[$\rho_{xx}^2$+$\rho_{xy}^2$], where $\rho_{xy}$ is the Hall resistivity. We have fitted the magneto-conductivity data with the HLN equation for fields below 2 T and shown in Fig. 2(b). The excellent fitting between experimental data and theoretical expression confirms that the low-temperature MR is due to the WAL effect. As shown in  Fig. 2(c), $l_{\phi}$ reduces drastically with increasing temperature, following the theoretically predicted temperature dependence, $l_{\phi}$$\propto$$T^{-0.5}$ \cite{alts,lee1,lee2,choi}. This observation further supports that the low-temperature transport is dominated by the surface state. We have also extracted the value of the parameter $\alpha$, which is $\sim$1 at  5 K and 10 K but it gradually decreases with increase in temperature and drops down to 0.8 at 40 K.  The value of $\alpha$ $\sim$1 implies that two conducting channels, top and bottom surfaces of the  single crystal, are taking part in charge conduction at low temperature.  At high temperatures, as significant fraction of the total current passes through the bulk, the value of $\alpha$ decreases from 1.\\
\begin{figure*}
\includegraphics[width=\textwidth]{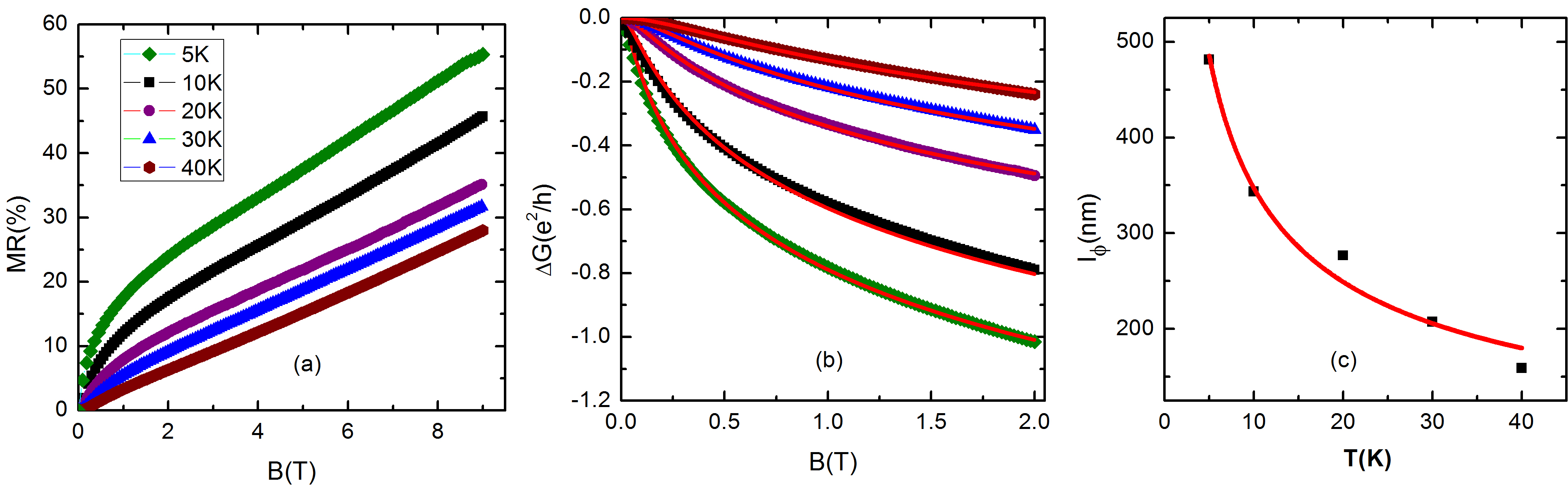}
\caption{(a) Magnetic field dependence of magnetoresistance (MR) of S2 crystal at different representative temperatures from 5 to 40 K. (b) The field dependence of surface conductance ($G$), which has been calculated using the the expression, $G$=$\frac{\rho_{xx}}{\rho_{xx}^{2}+\rho_{xy}^{2}}$$t$. Here, $\rho_{xy}$ is the Hall resistivity and $t$ is the thickness of the sample. Solid lines are the fit to the experimental data with HLN equation. (c) Dephasing length is plotted against temperature, which follows the expected  $T^{-0.5}$ dependence, as shown by the red line.}\label{Fig.3}
\end{figure*}

\textbf{Hall resistivity and aging effect.} The temperature dependence of the Hall coefficient ($R_{H}$) for several freshly prepared and aged crystals have been measured. For these crystals,  $R_{H}$ is positive and shows activation behavior at high temperatures. The positive sign of $R_{H}$ implies that holes are the majority carriers at high temperatures. $R_{H}$($T$) for both freshly prepared and aged S2 crystals are shown in Fig. 3 as representative. As the Fermi level of BSTS lies in the gap of conduction band and valance band of the bulk,  $R_{H}$ exhibits activation behavior \cite{Arakane}. The value of activation energy deduced from the Hall data is about 39 meV which is very close to that calculated from resistivity. Valance band being closer to the Fermi level, $p$-type carrier dominates the charge conduction in the bulk.  The observed behavior of $R_{H}$($T$) in the range 125-300 K is qualitative similar to earlier reports on samples close to optimum composition \cite{Pan, Ren, Taskin}. Though the surface dominance over the bulk is observed at much lower temperature ($\sim$40 K) in $\rho_{xx}$, it starts at as high as 125 K for $R_{H}$. In a two-band model, due to the high weightage of  mobility than the density of carrier [7], the Hall response below 125 K is dominated by the high mobility carrier from the Dirac cone surface state.  The freshly cleaved sample has $p$-type surface state but the aged sample shows $n$-type carrier as indicated by the sign of $R_{H}$ at low temperature. It has been observed that exposure to atmosphere causes electron doping  on the surface, as a result, the Fermi level shifts above the Dirac point \cite{Taskin}. This type of aging effect is quite common in topological insulator  \cite{zhao}. Thus the crystals those were exposed to air for longer time, $R_{H}$  starts to decrease sharply below 125 K and becomes negative at low temperature. \\
\begin{figure}
\includegraphics[width=0.5\textwidth]{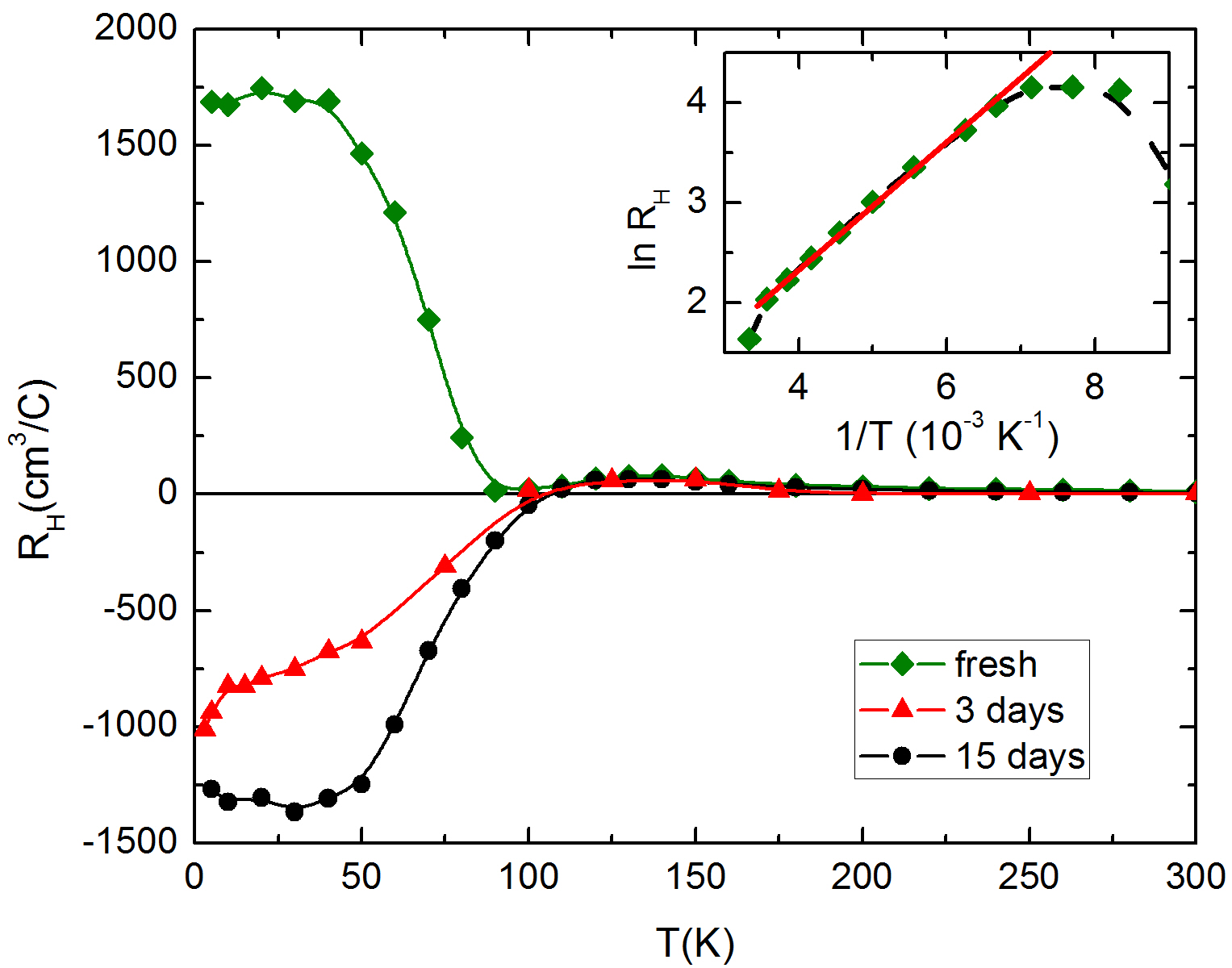}
\caption{Temperature dependence of the low-field Hall coefficient ($R_{H}$) for Bi$_{1.5}$Sb$_{0.5}$Te$_{1.7}$Se$_{1.3}$ single crystal (S2) at different ages. The blue dots are the $R_{H}$ data for the freshly cleaved S2 sample. The red and black dots represent $R_{H}$ data for the same S2 sample after being exposed to air for 3 days and 15 days respectively. Inset shows ln($R_{H}$) vs 1/$T$ plot for freshly cleaved S2 crystal. The solid line shows the linear behavior above 160 K. }\label{Fig.2}
\end{figure}

\textbf{Singular paramagnetic response in magnetization measurements.} Spin momentum locking is an integral characteristic of the surface state of a 3D topological insulator. As a result of strong spin-orbit coupling and time reversal symmetry, the spin and momentum wave vector of low-energy quasi particle excitations are always perpendicular to each other. This leads to left-handed spin texture for the conduction band and right-handed spin texture for the valence band on circular constant energy contour of the Dirac cones. However, at the Dirac node, where the two bands with opposite spin helicity touch each other, the spins are free to orient in any direction due to the singularity. As a consequence, it provides a paramagnetic contribution to the intrinsic magnetic moment of the system. Thus one expects to observe a low-field paramagnetic peak in the susceptibility curve $\chi$($B$). Experimental discovery of this singular peak in $\chi$ has been reported for the family of 3D topological insulators  and identified as the fingerprint of the helical spin texture of the  Dirac fermions of the surface state \cite{zhao,buga,pariari}. Also, $\chi$ shows linear-in-field decay from its zero-field value  \cite{zhao,buga,pariari}.  The linear field dependence of $\chi$ has  been established theoretically \cite{zhao}. Considering $T$=0, this paramagnetic Dirac susceptibility has the form,

\begin{equation}
\chi \cong \frac{\mu_{0}}{4\pi^{2}}[\frac{(g\mu_{B})^{2}}{\hbar v_{F}}\Lambda - \frac{(g\mu_{B})^{3}}{(\hbar v_{F})^{2}}\mid B\mid]
\end{equation}

at the sample's native chemical potential ($\mu$ = 0). Where $g$, $\mu_{B}$ and $v_{F}$ are the Land\'{e} $g$-factor, Bohr magneton and Fermi velocity respectively. $\Lambda$ is the effective size of the momentum space, which is responsible for the singular response in $\chi$. As the height of the peak in $\chi$ is determined by $\Lambda$, depending on the details of the bulk band, the peak height can vary from system to system within the family of 3D topological insulators \cite{zhao}. Whereas the nature of the topological response determined by the cuspiness, the linear-in-field decay, robustness against temperature, etc. of $\chi$($B$) are universal to the entire family of 3D topological insulators \cite{zhao}.\\
\begin{figure*}

\includegraphics[width=0.9\textwidth]{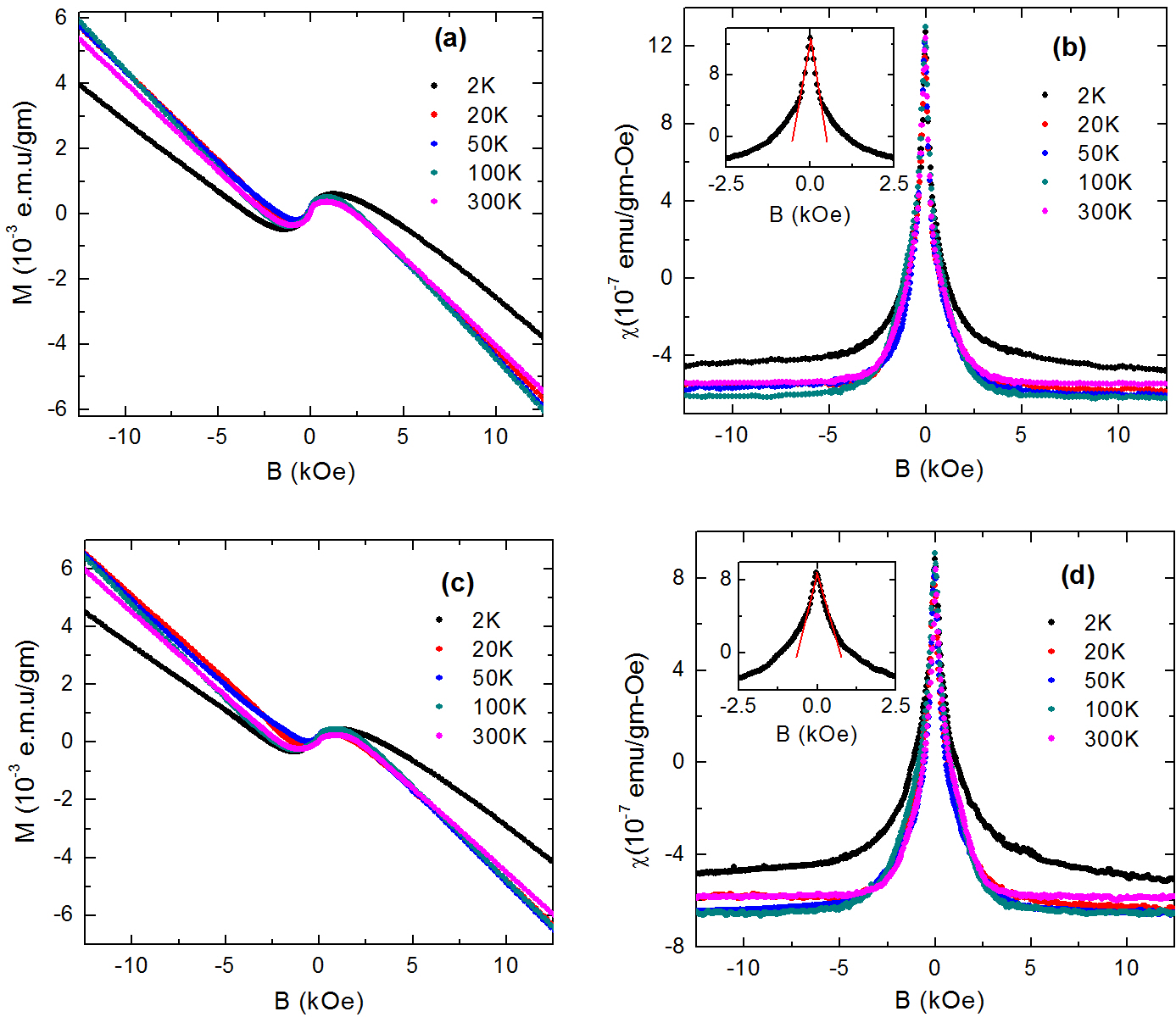}
\caption{(a) Magnetization ($M$) vs $B$ for freshly cleaved single crystal (S2) of Bi$_{1.5}$Sb$_{0.5}$Te$_{1.7}$Se$_{1.3}$ at several representative temperatures from 2 to 300 K, (b) Susceptibility \textbf{($\chi$=$\frac{dM}{dB}$)} as a function of $B$, calculated by taking the first order derivative of magnetization. Inset shows linear-in-field decay of $\chi$ from its zero field value at a representative temperature 2 K for the freshly cleaved sample. (c) Magnetization of the same piece of S2 single crystal of Bi$_{1.5}$Sb$_{0.5}$Te$_{1.7}$Se$_{1.3}$, which was kept for three days in air from the first measurement. (d) Susceptibility  as a function of $B$ at several representative temperatures. Inset shows linear-in-field decay of $\chi$ for aged sample at 2 K.}\label{Fig.4}
\end{figure*}

Figure 4(a) shows  magnetization ($M$) of the freshly cleaved single crystal (S2) as a function of magnetic field, at several representative temperatures between 2 and 300 K.  $M$ shows diamagnetic behaviour over a wide field range except at low fields, where a sharp paramagnetic upturn emerges. This is clearly visible from $\chi$ vs $B$ plot, as shown in Fig. 4(b). The experimental value of $\chi$ has been obtained after taking the first order derivative of $M(B)$ with respect to $B$. A cusp-like paramagnetic response in $\chi$($B$) sharply rises over the diamagnetic background in a narrow field range $\pm \sim$1.5 kOe. Irrespective of temperature, the height of the peak from the diamagnetic background remains almost same. This singular response shows no sign of thermal rounding up to the highest measuring temperature 300 K. Similar to Bi$_{2}$Se$_{3}$, Sb$_{2}$Te$_{3}$, Bi$_{2}$Te$_{3}$ \cite{zhao,buga} and ZrTe$_{5}$ \cite{pariari}, the paramagnetic response in the present system implies the helical spin texture of the 2D Dirac fermion on the surface. In this context, we would like to mention that the standard diamagnetic sample such as Bi does not show any  paramagnetic response in $\chi$ (see supplementary information). Inset of Fig. 4(b) shows linear decay of $\chi$ with field from its zero-field value at a representative temperature 2 K, as predicted theoretically (Eq. 2). When the surface is exposed to air for long time, the surface reconstruction and the formation of 2D electron gas occur due to the bending of bulk band   \cite{zhao}. As a result, the peak height has been observed to reduce with time \cite{zhao,pariari}. To probe the aging effect in BSTS, the magnetization measurements were done on the same piece of BSTS single crystal (S2) after 3 days of exposure to air and shown in Fig. 4(c). Although the nature of the  peak  along with the diamagnetic background, as shown in Fig. 4(d), remain unchanged, a significant drop ($\sim$ 25\%) is observed in the height, reflecting the expected aging effect in the present sample. Inset of Fig. 4(d)  shows the linear decay of $\chi$ with field at 2 K for the aged sample. \\

\textbf{\large Discussions}\\
We observe a crossover from insulating to metallic behaviour in the temperature dependence of resistivity in Bi$_{1.5}$Sb$_{0.5}$Te$_{1.7}$Se$_{1.3}$ single crystals, which can be associated with the dominating bulk and surface state contribution to transport respectively. The competition between surface and bulk conduction appears to be very sensitive to sample preparation.  A rare quadratic temperature dependent resistivity for metallic surface conduction has been observed for samples with large bulk resistivity. This highly decoupled  metallic surface, which was remained elusive in earlier studies, is an essential criterion for the use of spin-momentum locked topological state in electronic application and basic research.  In addition, we observe a sharp cusp-like paramagnetic peak at zero field in magnetization measurements. This temperature insensitive anomalous magnetic response,  has been associated with the helical spin texture of 2D Dirac fermion on the surface of 3D topological insulators. The surface state reconstruction with time due to doping from air, known as aging effect for 3D topological insulators, is clearly reflected from the age dependent reduction of the susceptibility peak.\\

\textbf{\large Method}\\
Single crystals of Bi$_{1.5}$Sb$_{0.5}$Te$_{1.7}$Se$_{1.3}$ were prepared by self-flux method using high purity elements (5N) of Bi, Sb, Te and Se in a stoichiometric ratio 1.5:0.5:1.7:1.3, respectively. Two slightly different techniques were adopted to prepare these crystals. In both the cases sample handling was done inside a glove box in argon gas atmosphere.  Method I:  Single crystal S3 was prepared in this method. Stoichiometric mixture of Bi, Sb, Te and Se  granules was vacuumed sealed in a quartz tube and then heated at the rate of 70$^{\circ}$C/h to 850 $^{\circ}$C and kept at that temperature for 48 h for diffusion process. The sample was cooled over a period of 60 h to 550 $^{\circ}$C and kept at that temperature for 96 h before furnace cooling.  Method II: Some minor modifications were done over method I to prepare single crystals S1 and S2. The stoichiometric mixture of elements was  pelletised before the vacuum sealing.  In this method, at the final stage (of method I) the sample was  slowly cooled from 550 $^{\circ}$C to room temperature at the rate of 10$^{\circ}$/h instead of furnace cooling. The local temperature gradient, which sustains due to the slow cooling process, in the vertically placed tube helps in stacking of chalcogenite layer and growing large single crystalline flake which can be easily cleaved  using a razor. The typical thickness of the samples used in electrical transport and magnetic measurements is $\sim$ 100 $\mu$m. In-plane electrical resistivity and Hall coefficient between 2 and 300 K were  measured by four-probe technique in a  physical property measurement system (Quantum Design). Magnetization measurement was performed in SQUID-VSM (Quantum Design). Before magnetization measurement of BSTS, the system was counterchecked by standard bismuth and palladium samples, the details of which are discussed in the supplementary  along with sample characterization.\\

\newpage
\pagebreak
\begin{center}
\large \textbf{Supplementary information for ``Prominent metallic surface conduction  and the singular magnetic response of  topological Dirac fermion in three-dimensional topological insulator Bi$_{1.5}$Sb$_{0.5}$Te$_{1.7}$Se$_{1.3}$"}
\end{center}
\subsection{\large Sample characterization using powder x-ray diffraction and energy dispersive x-ray techniques}
The phase purity of Bi$_{1.5}$Sb$_{0.5}$Te$_{1.7}$Se$_{1.3}$ (BSTS) samples was checked by high resolution powder x-ray diffraction (XRD) method with Cu K$_{\alpha}$ radiation in a Rigaku x-ray diffractometer (TTRAX II). Within the resolution of XRD, we have not observe any peak due to impurity phase. Fig. S5(a) shows the diffraction pattern for powdered crystal prepared by method II as a representative. The diffraction profile can be fitted well with hexagonal structure of space group symmetry R$\overline{3}$m using the Rietveld refinement. The calculated lattice parameters  are $a_{hex}$=4.2840 {\AA} and $c_{hex}$=29.871 {\AA}. Inset of Fig.S5(a) shows (1 0 7) and (0 0 12) peaks. The presence of these peaks are very important for ordering of chalcogen layer. We have also performed x-ray diffraction of a single flake as shown in Fig.S5(b). The Bragg peaks, which are integer multiple of (003) only are observed, indicating good $c$-axis orientation of the single crystal. The (0 0 12) peak is also very clearly observed in the diffraction pattern. Energy dispersive x-ray (EDX) (Quanta 200, FEG scanning electron microscope) analysis was used to determine the chemical composition of the sample. Identical chemical composition at different positions of  single flake and for different flakes suggest that the samples are chemically homogeneous. Fig.S6 shows the typical EDX spectrum of one such BSTS single crystal.  The chemical composition determined using EDX is close that of nominal composition, at least within the limits of the SEM-EDX analysis.\\

\begin{figure*}[h!]
\includegraphics[width=\textwidth]{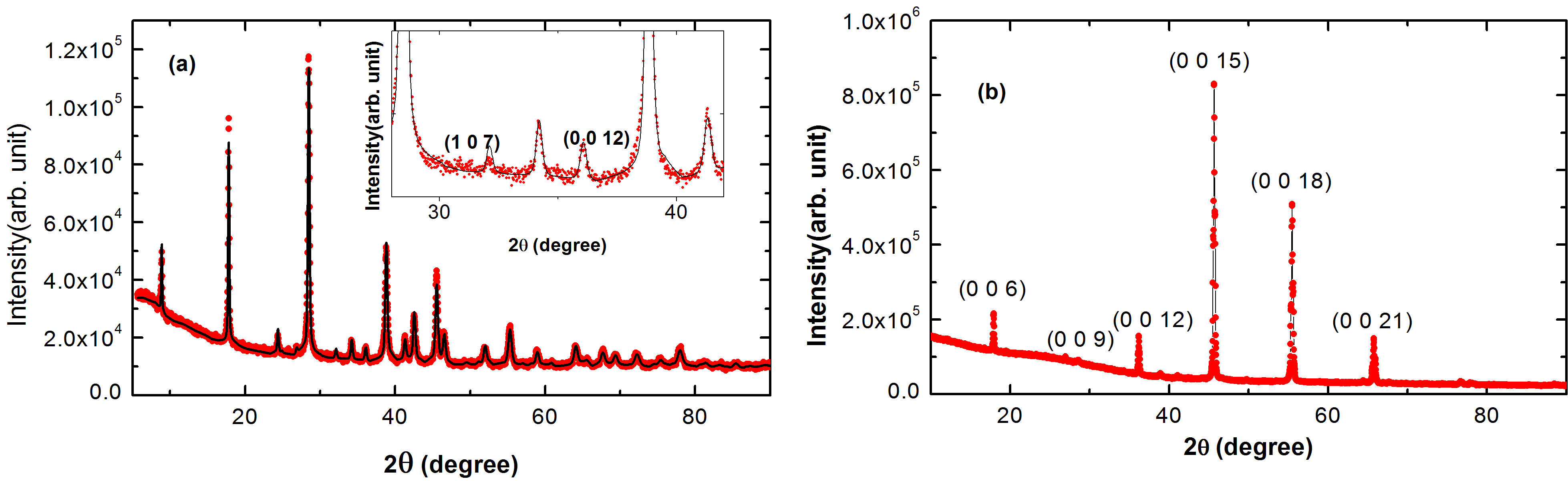}
\renewcommand{\figurename}{FIG.S}
\caption{(a) X-ray diffraction pattern of powdered sample of Bi$_{1.5}$Sb$_{0.5}$Te$_{1.7}$Se$_{1.3}$ single crystals. Red open circles are experimental data and the black continuous line corresponds to Rietveld refinement of the diffraction pattern. The Bragg peaks (1 0 7) and (0 0 12) are shown in the inset. (b) X-ray diffraction pattern for a single flake shows less number of Bragg peaks than the powdered sample with peak index integer multiple of (003).}\label{rh}
\end{figure*}

\begin{figure}
\includegraphics[width=0.7\textwidth]{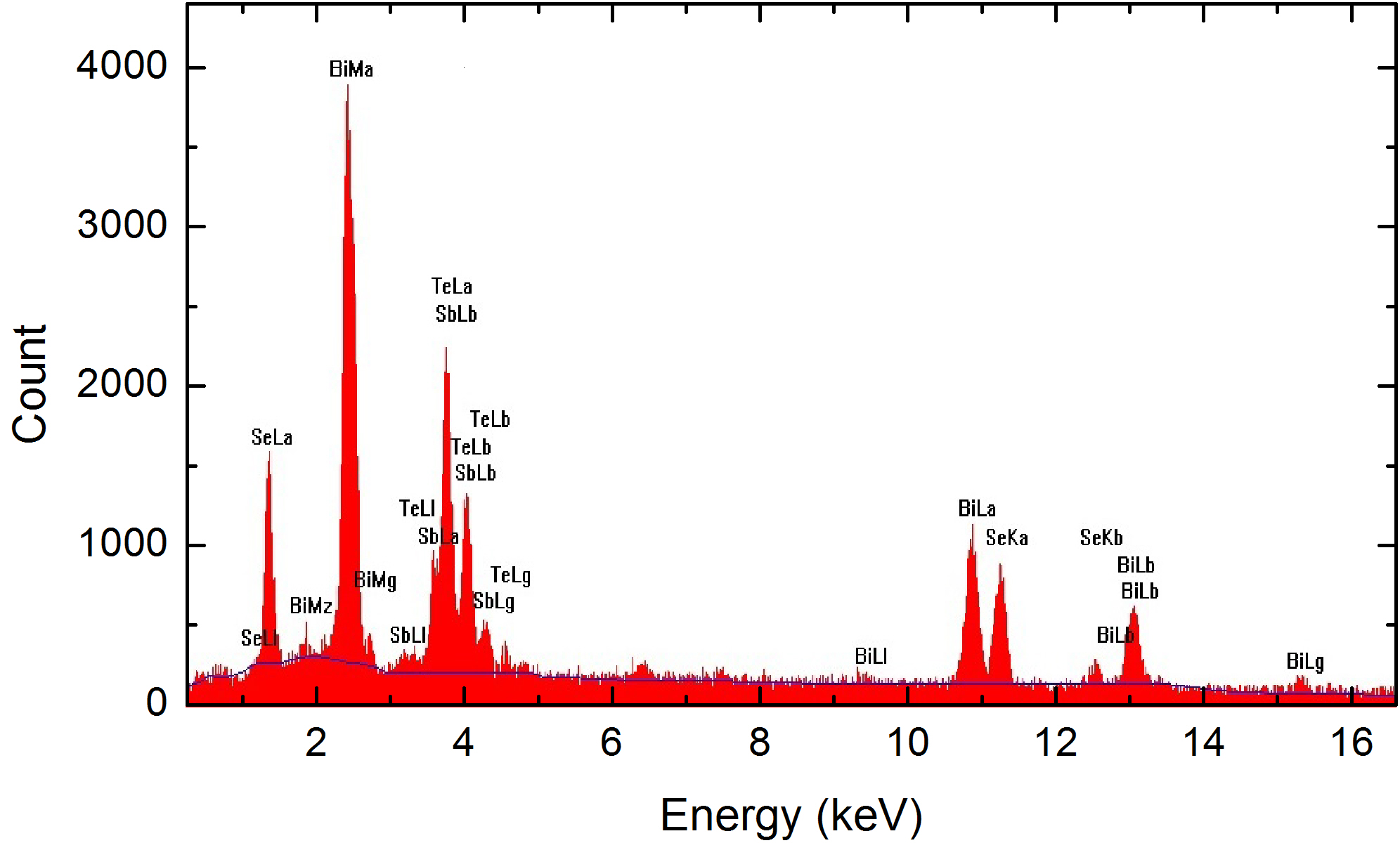}
\renewcommand{\figurename}{FIG.S}
\caption{Energy dispersive x-ray spectroscopy of Bi$_{1.5}$Sb$_{0.5}$Te$_{1.7}$Se$_{1.3}$ single crystal.}\label{edax}
\end{figure}
\pagebreak
\newpage
\subsection{\large Magnetic measurements of standard samples}
We have measured field response of magnetic moment of standard bismuth and palladium samples in SQUID-VSM [MPMS 3, Quantum Design] prior to BSTS single crystal. FIG.S7 (a) shows that linear diamagnetic moment of bismuth at 2 and 100 K passes through the origin. This is more clear from FIG.S7 (b), which shows magnetic field dependence of differential susceptibility ($\chi$=$\frac{dM}{dB}$). Absence of any paramagnetic peak around zero field implies that singular paramagnetic susceptibility in BSTS single crystal is not due to any spurious response in our system.\\

\begin{figure*}[h!]
\includegraphics[width=\textwidth]{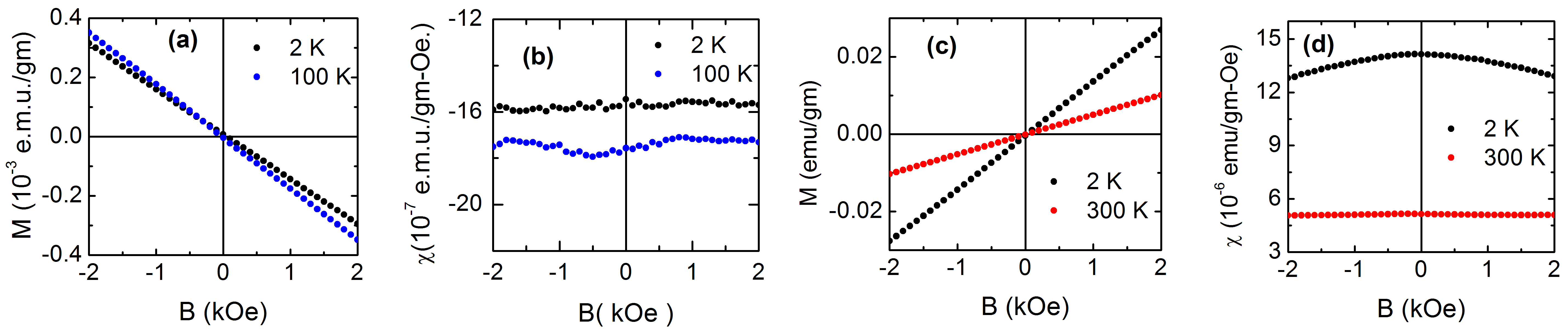}
\renewcommand{\figurename}{FIG.S}
\caption{(a) Magnetization measured  at several representative temperatures for standard diamagnetic bismuth sample. (b) Differential susceptibility ($\chi$=$\frac{dM}{dB}$) obtained after taking numerical derivative of the magnetization with respect to external magnetic field.
(c) Magnetization of standard paramagnetic palladium sample at 2 and 300 K. (d) Differential susceptibility ($\chi$=$\frac{dM}{dB}$) obtained after taking numerical derivative of the magnetization with respect to external magnetic field, }\label{rh}
\end{figure*}

FIG. S7(c) shows the expected magnetic behaviour of paramagnetic palladium sample, provided by  Quantum Design. FIG.S7(d) shows the low-field susceptibility of palladium at 2 K and room temperature. The nonlinear behaviour of $\chi$ at low field and a broad zero field peak at 2 K are completely suppressed with increasing temperature. This behavior is entirely different from the singular, robust and linear low-field paramagnetic response of the topological surface state in the present sample.\\

\subsection{\large Field dependence of the Hall resistivity}

The field dependence of Hall resistivity ($\rho_{xy}$) of freshly cleaved BSTS is plotted in Fig.S8 at some representative temperatures from 5 to 200 K. The slope of the $\rho_{xy}$ \emph{vs} $B$ plot at low field, provides the value of Hall coefficient (R$_{H}$) at that temperature as discussed in the manuscript. $\rho_{xy}$ is positive and increases almost linearly with field from room temperature to $\sim$ 120 K. This behavior of  R$_{H}$ indicates that $p$-type charge carrier dominates the conductivity of the bulk at high temperature. As the temperature decreases further, nonlinearity in $\rho_{xy}$ \emph{vs} $B$ starts to appear because the surface contribution becomes significant due to exponential decrease in bulk carrier. Thus, due to presence of two different bands with different carrier concentration and mobility, the field dependence of $\rho_{xy}$ becomes nonlinear, as shown in Fig.S8 at 5 and 50 K. Although $\rho_{xy}$ is weakly nonlinear at high fields, the low-field values of R$_{H}$ is almost same at low temperatures.
\begin{figure}[h!]
\includegraphics[width=0.5\textwidth]{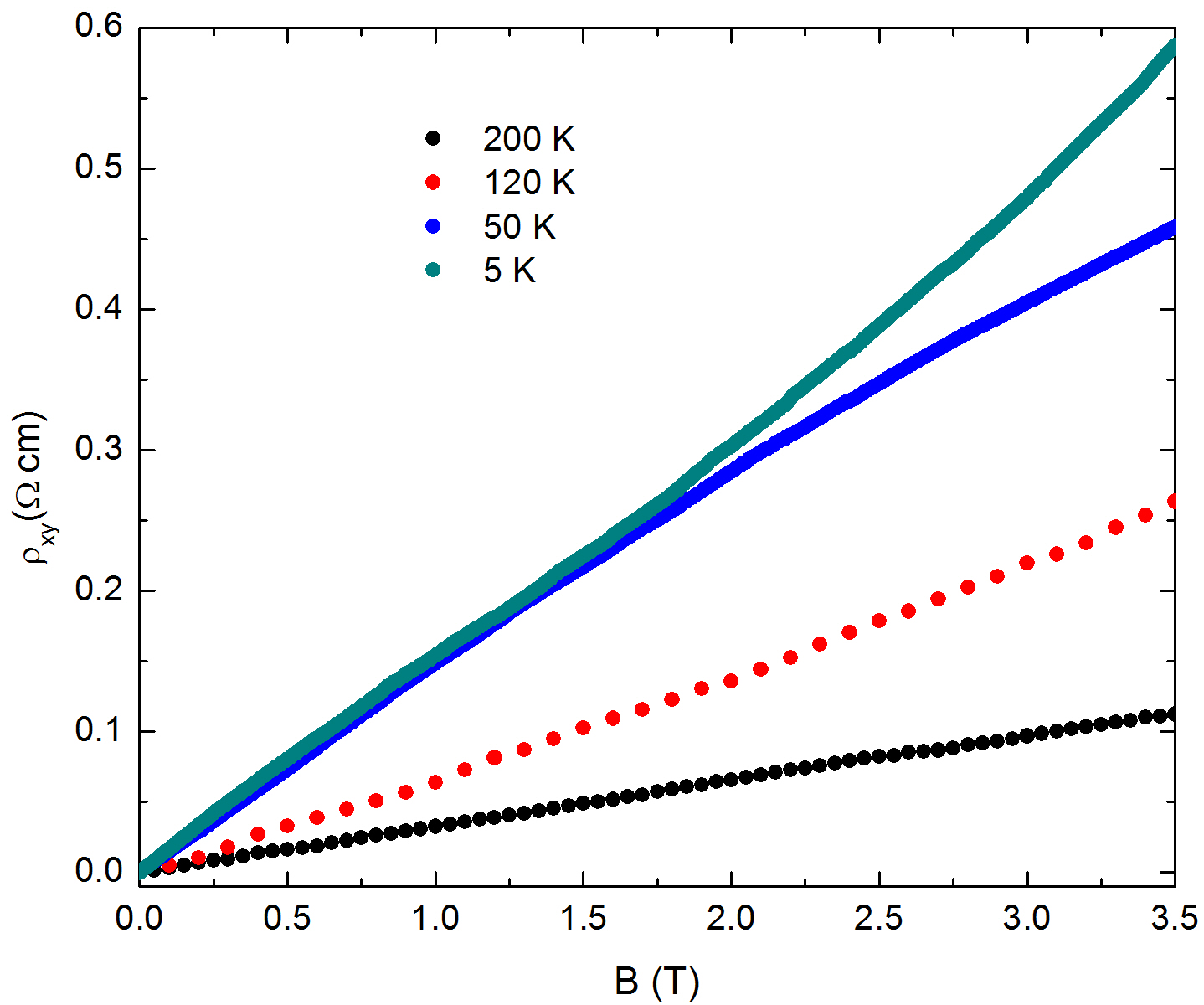}
\renewcommand{\figurename}{FIG.S}
\caption{Magnetic field dependence of Hall resistivity ($\rho_{xy}$) at some representative temperatures. $\rho_{xy}$ at 120 K and 200 K has been multiplied by factor 10.}\label{rhb}
\end{figure}
\end{document}